\documentclass[letterpaper,twocolumn,aps,prl,showpacs,amsmath,amssymb]{revtex4-1}

\usepackage{graphicx}
\usepackage{dcolumn}
\usepackage{bm}

\begin{document}

\title{Simulations of Coulombic Fission of Charged Inviscid
Drops}

\author{J.C. Burton}
\email{jcburton@uchicago.edu}
\affiliation{James Franck Institute and Department of Physics, The University of Chicago}
\author{P. Taborek}
\affiliation{Department of Physics and Astronomy, University of
California, Irvine}

\date{\today}

\begin{abstract} 
We present boundary-integral simulations of the evolution of critically charged droplets.  For such droplets, small ellipsoidal perturbations are unstable and eventually lead to the formation of a ``lemon"-shaped drop with very sharp tips.  For perfectly conducting drops, the tip forms a self-similar cone shape with a subtended angle identical to that of a Taylor cone. At the tip, quantities such and pressure and fluid velocity diverge in time with power-law scaling. In contrast, when charge transport is described by a finite conductivity, we find that small progeny drops are formed at the tips whose size decreases as the conductivity is increased. These small progeny drops are of nearly critical charge, and are precursors to the emission of a sustained flow of liquid from the tips as observed in experiments of isolated charged drops.
\end{abstract}

\pacs{47.65.-d, 47.15.km, 47.55.D-, 47.11.Hj}

\maketitle

An isolated droplet of liquid will naturally take the form of a sphere in order to minimize its surface area. If we now place a net amount of electrical charge on the sphere, there is a pressure opposing the effects of surface tension due to the repulsion of mutual charges. As first described by Lord Rayleigh \cite{rayleigh-PM1882}, there is a critical amount of charge $Q_{c}$ that can be placed on the drop before the sphere will become unstable to small perturbations from equilibrium: 
\begin{align} Q_{c}=8\pi(\epsilon\sigma R^{3})^{1/2} 
\label{critcharge} 
\end{align} 
where $R$ is the radius of the drop, $\sigma$ is the surface tension, and $\epsilon$ is the electrical permittivity. The simplest case, and the one first directly observed in experiment \cite{duft-NAT2003,achtzehn-EPJD2005}, is that of an ellipsoidal perturbation where the droplet evolves into a ``lemon" shape, and high-speed jets of liquid carrying a significant fraction of the total charge are emitted from the tips.

The disintegration of such isolated charged drops occurs in natural settings such as thunderstorm clouds and bursting bubbles at the ocean surface \cite{blanchard-JMET1958}, industrial applications ranging from ink-jet printing to electrospraying \cite{calvo-PRE2009}, and is especially important in mass spectrometry \cite{kebarle-JMS2000,grimm-JPHCH2005}. Charged liquid drops were also used as an early model for the mechanism of nuclear fission \cite{bohr-PR1939}. Nearly all previous studies have looked solely at the oscillations around equilibrium and the limits of stability \cite{swiatecki-PR1956,tsamopoulos-PRSL1985,basaran-PFA1989}, or operated under the assumption of infinite conductivity \cite{fontelos-POF2008,smith-PRSL1971,giglio-PRE2008}. This latter assumption is especially important; recent numerical simulations show that bulk conductivity controls the fine fluid jets formed in applied electric fields \cite{collins-NATPHYS2007}. Indeed, the specific mode of charge conduction will affect the dynamics of the jet \cite{mora-ARFM2007}. The most popular models of charge transport suppose bulk conduction \cite{saville-ARFM1997,collins-NATPHYS2007}, although recent evidence suggests that conduction along the surface of the drop is significant \cite{hunter-PCCP2009,burcham-JFM2002}.

In this letter, we simulate the initial instability and eventual cone-jet formation for charged inviscid drops with total charge $Q_{c}$. We consider two separate cases: (1) The drop is a perfect conductor. In this regime the ``lemon"-shaped drop forms extremely sharp tips, where quantities such as charge density, curvature, and velocity diverge in finite time. The tip shape is self-similar and conical, and the subtended angle is exactly that of a Taylor cone, which results from the curvature term being present in the asymptotic balance of forces. (2) The charge transport is limited by a finite electrical conductivity. We investigate the effects of bulk and surface conduction of charge; in both cases the limited conductivity results in a blunting of the sharp cone-tip, and the eventual emission of a progeny drop from the tip. Regardless of the method of charge transport, the progeny drop carries an amount of charge just below the Rayleigh limit $Q_{c}$, which depends weakly on its size. 
\begin{figure} \begin{center}
\includegraphics[scale=.9]{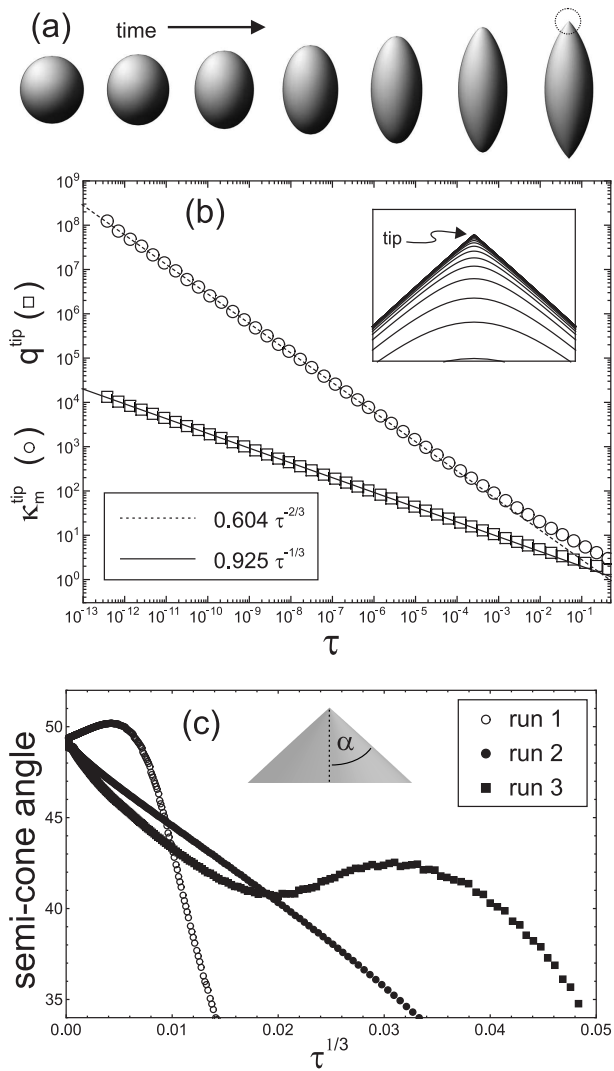}
\caption[]{$\textbf{(a)}$ Evolution of the drop from a slightly prolate shape to the final ``lemon" shape with pointed cone tips. $\textbf{(b)}$ Mean curvature ($\kappa^{tip}_{min}$) and charge density ($q^{tip}$) at the tip as a function of $\tau$ for an infinitely conducting droplet in a vacuum. The tip region (dotted circle in (a)) seen in the inset shows the evolution of the droplet interface into a Taylor cone. The quantities $\kappa^{tip}_{min}$ and $q^{tip}$ scale as $\tau^{-2/3}$ and $\tau^{-1/3}$, respectively. The solid and dashed lines show power-law fits to the data with fixed exponents. $\textbf{(c)}$ Time evolution of the semi-cone angle for 3 different simulations. The semi-cone angle always asymptotically approaches Taylor's value of $\approx$ 49.3$^{\circ}$, regardless of initial conditions, indicating a universal self-similar solution. The approach of the semi-cone angle scales roughly as $\tau^{1/3}$, thus the choice of abscissa.} 
\label{scalingplot}
\end{center} 
\end{figure}

\textit{Numerical method}---We begin with an incompressible, axisymmetric fluid globule immersed in an incompressible fluid of infinite extent. The axis of rotation is the $z$-axis and $r$ is the radial coordinate. The inner fluid has a density $\rho_{1}$ and the exterior fluid has a density $\rho_{2}$, where the density ratio $\Lambda=\rho_{2}/\rho_{1}$.  With this formulation we can consider the effects of an exterior fluid, so that the behavior of both bubbles and droplets can be studied, although in this letter we will focus solely on the case where $\Lambda\leq0.001$. The interface also has a uniform surface tension $\sigma$ and a varying surface charge density $q$ with total charge $Q_{c}$.  The charge conduction along the surface can be perfect so that $q$ only depends on the geometry of the drop, or transport properties can be specified with a bulk conductivity $k$ or surface conductivity $\gamma$. The flow everywhere is assumed to be inviscid and irrotational (smooth and non-turbulent). With these assumptions the velocity of the fluid $\vec{\textbf{v}}$ can be described by the gradient of a scalar potential $\vec{\textbf{v}}=\vec{\nabla}\phi$, and the problem is reduced to solving Laplace's equation $\vec{\nabla}^{2}\phi=0$ with time-dependent boundary conditions \cite{lister-PF2003,giglio-PRE2008}:
\begin{align} 
\label{kinematicbc1}
\left(\frac{\partial\phi}{\partial
t}+\frac{\left|\vec{\textbf{v}}\right|^{2}}{2}\right)^{-}-\Lambda\left(\frac{\partial\phi}{\partial
t}+\frac{\left|\vec{\textbf{v}}\right|^{2}}{2}\right)^{+}=
2q^{2}-2\kappa_{m}
\end{align} 
where $\kappa_{m}$ is the mean curvature and the superscripts refer to the exterior (+) or interior (-) side of the interface. In eqn.\ \ref{kinematicbc1} and all further discussions, quantities have been made dimensionless with lengths scaled by the initial radius $R$ of the globule. All times in the problem are scaled by ($R^{3}\rho_{1}/\sigma$), and any units of charge are scaled with 2($\epsilon\sigma R^{3}$)$^{1/2}$ so that critical surface charge density is equal to unity. For simplicity, we also assume the electrical permittivity everywhere is equal to $\epsilon$. 

We follow previous boundary-integral methods \cite{lister-PF2003,monika-JCP2004,burton-POF2007} used to accurately compute the motion of an interface between two inviscid fluids, with addition of the charge density in eqn.\ \ref{kinematicbc1}.  In our formulation there is no bulk free charge and the total charge on the interface is conserved. If the conductivity is infinite, then the charge distribution is purely geometry-dependent and the charge will always arrange itself so that the electric potential $\psi$ is constant.  In this case we solve the fairly simple electrostatics problem of finding the surface charge distribution on a charged axisymmetric conductor, which involves inverting an integral equation.  

However, this is a special case. When the conductivity is finite, we follow the method described in reference \cite{stone-POFA1990}, with the addition of terms accounting for Ohmic conduction \cite{saville-ARFM1997,collins-NATPHYS2007}:
\begin{align} 
\label{kinematicbc2}
\frac{Dq}{Dt}=\frac{1}{Pe}\nabla_{s}^{2}q-2q\kappa_{m}v_{n}+\Gamma\nabla_{s}^{2}\psi+K(\vec{\textbf{n}}\cdot\vec{\nabla}\psi)^{-}
\end{align} 
where $v_{n}$ is the normal velocity of the interface, $\nabla_{s}$ is the surface gradient operator, and the convective derivative operator is $D/Dt=\partial/\partial t+v_{n}\vec{\textbf{n}}\cdot\vec{\nabla}$. The dimensionless number $Pe=(R\sigma/\chi^{2}\rho)^{1/2}$ is the P$\acute{e}$clet number, where $\chi$ is the surface diffusivity. For all simulations, we use a value of $Pe$=1000, so that diffusion of charge is essentially negligible. The parameters $K=k(R^{3}\rho/\sigma\epsilon^{2})^{1/2}$ and $\Gamma=\gamma(R\rho/\sigma\epsilon^{2})^{1/2}$ are a measure of the bulk and surface conductivity, respectively, and each is defined as a ratio of time scales, $t_{conduction}/t_{capillary}$, where the conduction timescale depends on the mode of conduction, bulk or surface. For our simulations, we varied $K$ and $\Gamma$ independently while the other was set to zero so we could isolate the effects of a particular transport coefficient.  
\begin{figure} 
\begin{center}
\includegraphics[scale=.88]{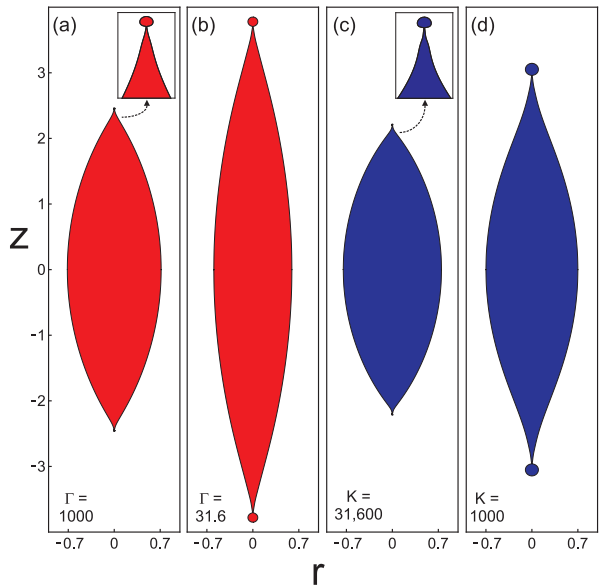}
\caption[]{Final drop shapes at the time of progeny drop emission for several values of the dimensionless surface conductivity $\Gamma$ and bulk conductivity $K$. The drop is surrounded by dilute vapor ($\Lambda$=0.001). The initial evolution of the shapes is similar to the infinite conductivity case (Fig.\ \ref{scalingplot}(a)) with the addition of a progeny drop formed at the tips. Lower values of $\Gamma$ as seen in (b), or $K$ as seen in (d), lead to more elongated drops and larger progeny drops because the charge moves more slowly.  The insets are zoomed-in images of the tip. The specific mode of charge conduction determines the shape of the drop and tip prior to emission.}
\label{drop_sequence}
\end{center} 
\end{figure}

\textit{Perfect Conductors}---First we will consider the case of a perfectly conducting drop.  For simplicity, we ignore the exterior fluid and choose $\Lambda$=0. The charge distribution is calculated purely based on the interfacial geometry at each time step. After the lemon-shape is formed (Fig.\ \ref{scalingplot}(a)), the tip continues to sharpen indefinitely: both the curvature and the charge density diverge. Fig.\ \ref{scalingplot}(b) shows the mean curvature $\kappa_{m}^{tip}$ and charge density $q^{tip}$ at the tip as a function of $\tau=t_{o}-t$, where $t_{o}$ is a parameter chosen from a power-law fit of the data, and represents the point of divergence. The mean curvature scales as $\tau^{-2/3}$, and the charge density as $\tau^{-1/3}$. Although not included in Fig.\ \ref{scalingplot}(b), we find that the normal velocity of the interface at the tip scales as $\tau^{-1/3}$. These scalings indicate that all terms in eqn.\ \ref{kinematicbc1} will be important for an asymptotic balance. This is in contrast to the Coulombic fission of a viscous, perfectly conducting droplet, where the charge density scales as $\tau^{-0.5}$ and the curvature as $\tau^{-0.72}$ \cite{fontelos-POF2008}, so that the curvature term becomes negligible as $\tau\rightarrow0$ and the viscous forces balance the electrostatic forces. Similar exponents were also measured experimentally in \cite{oddershede-PRL2000}, which considered the formation of a spout from an oil-water interface in an applied electric field.

The power-law behavior in the dynamics suggest a self-similar solution for the tip region. Our results show that the asymptotic shape of the tip region is that of a perfect cone, in contrast to simulations in reference \cite{giglio-PRE2008} with a much lower dynamic range in space and time. Surprisingly, the semi-cone angle is exactly equal to that of a Taylor cone, $\alpha_{T}\approx$ 49.3$^{\circ}$. Why is this so? Taylor \cite{taylor-PRSA1964} assumed a $\textit{steady-state}$ solution where the electrostatic pressure exactly balanced the curvature pressure on the surface of a cone (last 2 terms in eqn.\ \ref{kinematicbc1}). He showed that $V\propto r_{a}^{1/2}P_{1/2}(\cos\theta)$ in the far field, where $r_{a}$ is the distance from the apex of the cone, $P$ is a Legendre polynomial,  and $\theta=\pi-\alpha$ is the obtuse angle to the vertical (Fig.\ \ref{scalingplot}(c)). For a perfect conductor $\psi$ is constant, so $P_{1/2}(\cos\theta)=0$ and $\theta=\pi-\alpha_{T}$. For our dynamic case, one realizes that the velocity potential also satisfies Laplace's equation, and $\phi$ should be of the same form as $\psi$. Thus capillary, electric, and Bernoulli forces all balance in the asymptotic cone shape, and the cone angle must be $\alpha_{T}$. This solution is apparently universal, as different initial simulation conditions all converge to the same value for the cone angle in Fig.\ \ref{scalingplot}(c).

\textit{Finite Conductivity}---Next we will consider the case where the charge transport is limited by a finite electrical conductivity. For these simulations we use a density ratio $\Lambda$=0.001, corresponding to a liquid drop in an ambient gas atmosphere. Previous experimental studies of charged drops \cite{hunter-PCCP2009} and liquid bridges \cite{burcham-JFM2002} suggest that both bulk and surface conduction are important in ionic solutions due to a diffuse layer near the surface of the liquid. We have simulated both cases independently to elucidate the effects that the mode of charge transport has on the drop dynamics. Fig.\ \ref{drop_sequence} shows the final shapes of the drop just before progeny drop emission for two values of $\Gamma$ (a)-(b), and for two values of $K$ (c)-(d). In both cases, for large conductivities, the final shape resembles the perfect conductor, with the addition of a tiny drop emitted from the tip. Our simulations cannot proceed past the emission of the first drop, after which a jet/stream of droplets have been visualized in experiments \cite{duft-NAT2003,giglio-PRE2008}. Lower conductivities result in large progeny drops and highly elongated drop shapes, which is due to the reduced rate of charge transport to the pointed tips. As the tip sharpens, the charge must move to the tip region to remain in equilibrium. If this cannot occur quickly enough, then the surface tension forces will act to breakup elongated structures through the Rayleigh-Plateau instability. At the final moments of progeny drop pinch-off, the charge density remains finite and the dynamics are governed by the known universal solution for inviscid capillary pinch-off \cite{lister-PRL1998}.
\begin{figure} 
\begin{center}
\includegraphics[scale=.88]{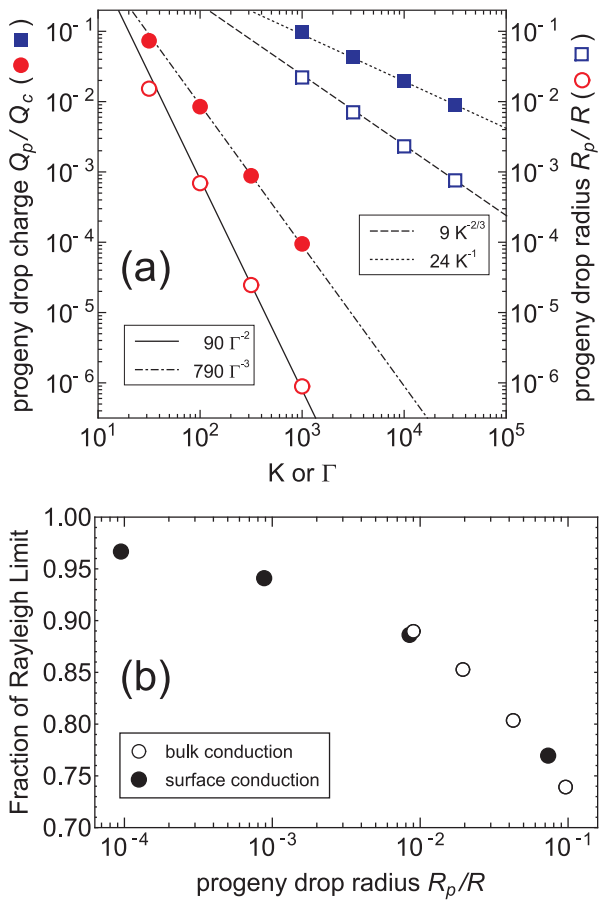}
\caption[]{$\textbf{(a)}$ Radius of the progeny drop $R_{p}$ (closed symbols) and the total charge on the progeny drop $Q_{p}$ (open symbols) at the moment of pinch-off. Red circles refer to simulations using surface conduction and blue squares using bulk conduction. The solid and dot-dashed lines show a power-law fit of $\Gamma^{-3}$ and $\Gamma^{-2}$, respectively. The dashed and dotted lines show power-law fits of $K^{-1}$ and $K^{-2/3}$. These scaling laws are expected from equating the conduction and capillary time scales for surface and bulk conduction. As the conductivity is increased, the size of the progeny drops become smaller and the total charge they carry decreases. $\textbf{(b)}$ Ratio of the total charge on the progeny drop to the critical amount of charge necessary for the instability (eqn.\ \ref{critcharge}). This is the fraction of the Rayleigh limit, so that a value of 1 means the progeny drop will certainly be unstable and undergo further fission. Regardless of the conduction mechanism, this fraction depends only on the drop size.}
\label{radius_charge_plot} 
\end{center} 
\end{figure}

Fig.\ \ref{radius_charge_plot}(a) shows the dependence of the progeny drop radius $R_{p}$ and charge $Q_{p}$ on the conductivity parameters $\Gamma$ and $K$. For bulk conduction, $R_{p}\propto K^{-2/3}$ and $Q_{p}\propto K^{-1}$. The progeny drop size can be understood as a balance of the conduction time scale $t_{conduction}=\epsilon/k$ with the local capillary time scale $t_{capillary}=(R_{p}^{3}\rho/\sigma)^{1/2}$. If we assume that the progeny drop has of order the Rayleigh limit of charge (eqn.\ \ref{critcharge}), then we obtain the $K^{-1}$ scaling for $Q_{p}$ as well.  This argument is identical for the case of surface conduction where  $R_{p}\propto \Gamma^{-2}$ and $Q_{p}\propto\Gamma^{-3}$, except the conduction time scale is now defined by the surface conductivity $t_{conduction}=\epsilon R_{p}/\gamma$. The observed scalings in the simulations are in excellent agreement with these predictions. Fig.\ \ref{radius_charge_plot}(b) shows the ratio of $Q_{p}$ to the Rayleigh limit for the progeny drop $8\pi(\epsilon\sigma R_{p}^{3})^{1/2}$. Regardless of the charge conduction mechanism, this ratio only depends on the progeny drop radius $R_{p}$, and slowly approaches unity for small drops. This suggests that drops emitted from the tips during a Coulombic fission process are only marginally stable, and could be subject to further breakup given a sufficient perturbation. 

\textit{Conclusion}---For the first time, we provide a quantitative picture which shows how charge conduction controls the shape and eventual emission of drops during the Coulombic fission of inviscid drops. For perfectly conducting drops, a self-similar cone shape is formed with a cone angle of $\approx$ 49.3$^{\circ}$, identical to Taylor's steady-state value. Although our simulations are dynamic and the pressure and velocity are diverging at the tip of the cone, this cone angle can be understood solely by the presence of the mean curvature in the asymptotic balance of forces. When the charge transport is limited by a finite surface or bulk conductivity, a small progeny drop is emitted from the tip of the cone; a precursor to the tip-jetting observed in experiments. The drop size and charge are determined by a ratio of the conduction time scale to the capillary time scale, and the progeny drop is nearly unstable and may undergo further Coulomb fission. 

We are grateful to Sidney Nagel for helpful discussions.  This work was supported by the ICAM post-doctoral fellowship program and NSF grant DMR 0907495.

\end{document}